\documentclass[twocolumn,aps,prc,floatfix]{revtex4}
\usepackage{graphicx}
\usepackage{dcolumn}
\usepackage{bm}

\begin{document}

\title{Probing proton transition momentum in neutron-rich matter}

\author{Gao-Chan Yong}

\affiliation{%
Institute of Modern Physics, Chinese Academy of Sciences, Lanzhou
730000, China
}%

\begin{abstract}

Around the nuclear Fermi momentum, there is a transition of nucleon momentum distribution n(k) in nuclear matter, i.e., from a constant to the $1/k^{4}$ nucleon momentum distribution. While nowadays the transition momentum of minority in asymmetric matter is rarely studied and thus undetermined. In the framework of the IBUU transport model, proton transition momentum in nuclei is first studied. It is found that the transition momentum of proton is sensitive to the $\pi^{-}/\pi^{+}$ ratio as well as the energetic photon production in neutron-rich nuclear reaction. This result may push the study of how the proton momentum is distributed in neutron-rich matter forward and help us to better understand the dynamics of both neutron-rich nuclear reactions and neutron stars.

\end{abstract}

\maketitle

In recent two years, the study of nuclear short-range correlations attracts much attention \cite{RMP2017,caili4,yongp3,yongp4,yongprc2016,chenx17}.
It has been shown that about 20\%
nucleons in nuclei are correlated \cite{e93,e96,sci08}. Because of the nucleon short-range interactions \cite{tenf05,tenf07}, nucleons in nuclei can form pairs with larger
relative momenta and smaller center-of-mass momenta
\cite{pia06,sh07}. The nucleon short-range correlations (SRC) in nuclei cause a high-momentum tail (HMT) in single-nucleon momentum distribution above the Fermi-momentum
\cite{bethe71,anto88,Rios09,yin13,Claudio15}. And the HMT shape caused by two-nucleon SRC is almost identical for all nuclei from deuteron to very heavier nuclei \cite{Ciofi96,Fantoni84,Pieper92,egiyan03}, i.e., roughly exhibits
a $1/k^{4}$ tail \cite{hen14,sci14,henprc15,liba15}.
And in the HMT, the number of neutron-proton correlated pairs is about 18 times that of the proton-proton or neutron-neutron correlated pairs \cite{sci08}.

Proton transition momentum, i.e., the starting point of proton $1/k^{4}$ momentum distribution in neutron-rich matter directly affects proton average kinetic energy in nuclear matter, thus affects the dynamics of neutron-rich nuclear reactions and the dynamics of neutron stars, such as the cooling of a Neutron Star, the superfluidity
of protons \cite{mm2013}, etc. While for neutron-rich matter, it is not straightforward to determine the transition momentum of proton.
One general considers that below the Fermi momentum, proton or neutron have independent movements while above their respective Fermi momenta, i.e., $k_{F_{p}}$, $k_{F_{n}}$, they respectively have $1/k^{4}$ distributions starting from their respective Fermi momenta. This naive opinion, however, is not consistent with the correlation picture of neutron-proton pair \cite{sci14}. The correlated neutron and proton should have almost the same momentum whether in symmetric or in asymmetric matter. In asymmetric matter or neutron matter, such as the neutron stars, neutron and proton may have very different Fermi momenta. If each correlated neutron and proton have $1/k^{4}$ distributions starting from their respective Fermi momenta, then the correlated neutron and proton would have very different momenta. This point evidently contradicts the thought of the n-p dominance model \cite{sci14}.

In neutron-rich matter, because protons become more prominent at high momenta as their concentration decreases \cite{Rios14}, the starting momentum of minority $1/k^{4}$ distribution should be not the minority Fermi momentum. Because the minority Fermi momentum would become very small in magnitude if proton concentration decreases sharply.
Apart from its own Fermi momentum, the left case is using the majority Fermi momentum as the starting momentum of minority $1/k^{4}$ distribution. For very asymmetric nuclear matter, it is hard to obtain the minority transition momentum from microscopic theory. The ladder self-consistent Green function approach could not get the nucleon momentum distribution at Zero temperature \cite{Rios09} and the Brueckner theory with a microscopic Three-body force gives a noncontinuous nucleon momentum distribution \cite{wangp2013}. That is to say, the microscopic theory can not answer the question of minority transition momentum in asymmetric matter, especially in the neutron matter. However, the minority transition momentum could be checked by nuclear experiments with unequal numbers of proton and neutron.

One way to check the the starting point of proton $1/k^{4}$ distribution, i.e., whether the HMT starts from proton Fermi momentum or from the correlated majority Fermi momentum, is using heavy-ion collisions with neutron-rich nuclei. In nucleus-nucleus collisions at intermediate energies, different proton energies may cause the difference of meson or photon productions in the final stage. In this study, it is found that charged pion ratio or hard photon production in neutron-rich nuclear reactions are sensitive to the correlated proton momentum, thus can be used to probe the starting point of proton $1/k^{4}$ distribution.

\begin{figure}[th]
\centering
\includegraphics[width=0.5\textwidth]{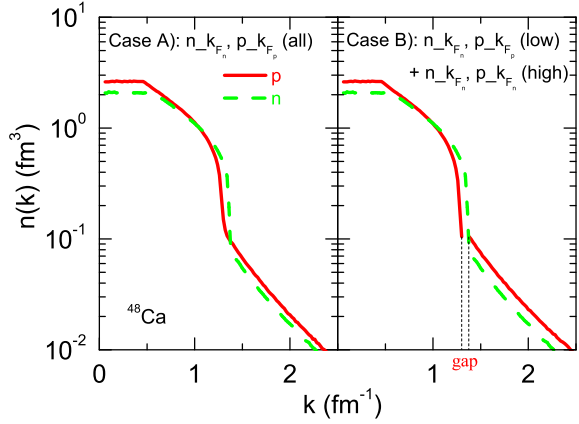}
\caption{ (Color online) Momentum distribution n(k) of nucleon with different starting points of proton $1/k^{4}$ distribution in nucleus $^{48}$Ca with normalization condition $\int_{0}^{k_{max}}n(k)k^{2}dk$ = 1. Case A stands for the HMT starting point of minority proton is its own Fermi-momentum while Case B starts the HMT of proton from majority neutron Fermi-momentum.} \label{npdis}
\end{figure}
In the used Isospin-dependent
Boltzmann-Uehling-Uhlenbeck (IBUU) transport model, neutron and proton initial density distributions in nuclei are given by the Skyrme-Hartree-Fock with Skyrme
M$^{\ast}$ force parameters \cite{skyrme86}. Nucleon
momentum distribution with high-momentum tail reaching about 2.75 times local Fermi momentum is adopted \cite{henprc15}. Since for medium and heavy nuclei there is a rough 20\% depletion of nucleon distributed in the Fermi sea \cite{sci08,sci14}, I let nucleon momentum distributions in nuclear matter piece of nuclei
\begin{eqnarray}
n(k)=\left\{%
  \begin{array}{ll}
    C_{1}, & \hbox{$k \leq k_{F}$;} \\
    C_{2}/k^{4}, & \hbox{$k_{F} < k < k_{max}$} \\
\end{array}%
\right.
\label{nk}
\end{eqnarray}
with normalization condition
\begin{equation}
\int_{0}^{k_{max}}n(k)k^{2}dk = 1
\end{equation}
and keep 20\% fraction of total nucleons in the HMT
\begin{equation}
\int_{k_{F}}^{k_{max}}n^{HMT}(k)k^{2}dk \bigg/ \int_{0}^{k_{max}}n(k)k^{2}dk = 20\%.
\end{equation}
In the n-p dominance picture \cite{sci14,sargnp14}, one needs to keep the same numbers of neutrons and protons in the HMT, thus
\begin{equation}
n^{HMT}_{p}(k)/n^{HMT}_{n}(k)= \rho_{n}/\rho_{p}.
\end{equation}
In the above equations, $k_{F}$ is nuclear Fermi momentum. $\delta$ denotes the local asymmetry $(\rho_{n}-\rho_{p})/\rho$, $\rho_{n}$ and $\rho_{p}$ are, respectively, local neutron and proton densities. The parameters $C_{1}$ and $C_{2}$ in Eq.~(\ref{nk}) can be automatically determined from the above equations.
The n-p dominance picture causes the inverse proportionality of the strength of the high-momentum distribution of protons and neutrons in neutron-rich matter, i.e., compared with majority neutrons, minority protons have larger probability with momenta greater than the Fermi momentum \cite{sargnp14}. This phenomenon has been confirmed by the recent experiments \cite{sci14}.
By using the local Thomas-Fermi relation
\begin{equation}
 k_{F_{n,p}}(r)= [3\pi^{2}\hbar^{3}\rho(r)_{n,p}]^{\frac{1}{3}},
\end{equation}
nucleon momentum distribution in nuclei is given by
\begin{equation}
 n_{n,p}(k)= \frac{1}{N,Z}\int _{0}^{r_{max}}d^{3}r\rho_{n,p}(r)\cdot n(k,k_{F}(r)),
\end{equation}
with $N$ and $Z$ being the total numbers of neutrons and protons in nuclei.
In Fig.~\ref{npdis}, nucleon momentum distribution of $^{48}$Ca is plotted. It is clearly seen that there is a HMT above the nuclear Fermi momentum. Proton has greater probability than neutron to have momenta greater than the nuclear Fermi momentum. While compared case A with case B, it is seen that with the starting point of majority Fermi momentum, proton has even more greater probability to have high momenta. This consequence may affect the dynamics of heavy-ion collisions at intermediate energies.

In the study, the isospin- and momentum-dependent single nucleon
potential is used \cite{spp1,yongprc2016}, i.e.,
\begin{eqnarray}
U(\rho,\delta,\vec{p},\tau)&=&A_u(x)\frac{\rho_{\tau'}}{\rho_0}+A_l(x)\frac{\rho_{\tau}}{\rho_0}\nonumber\\
& &+B(\frac{\rho}{\rho_0})^{\sigma}(1-x\delta^2)-8x\tau\frac{B}{\sigma+1}\frac{\rho^{\sigma-1}}{\rho_0^\sigma}\delta\rho_{\tau'}\nonumber\\
& &+\frac{2C_{\tau,\tau}}{\rho_0}\int
d^3\,\vec{p^{'}}\frac{f_\tau(\vec{r},\vec{p^{'}})}{1+(\vec{p}-\vec{p^{'}})^2/\Lambda^2}\nonumber\\
& &+\frac{2C_{\tau,\tau'}}{\rho_0}\int
d^3\,\vec{p^{'}}\frac{f_{\tau'}(\vec{r},\vec{p^{'}})}{1+(\vec{p}-\vec{p^{'}})^2/\Lambda^2},
\label{buupotential}
\end{eqnarray}
where $\rho_0$ denotes saturation density, $\tau, \tau'=1/2(-1/2)$ for neutron (proton).
And for nucleon-nucleon collisions, the isospin-dependent reduced nucleon-nucleon scattering cross section in medium is used. More details about the above single nucleon potential and baryon-baryon cross section can be found in Refs. \cite{yongp4,yongprc2016}.
In the model, the details on pion production through $NN\rightleftharpoons N\Delta$ and $\Delta\rightleftharpoons N\pi$ processes can be found in Ref. \cite{pion2017}.
The probability of energetic photon production from neutron-proton bremsstrahlung is given by
the one boson exchange model \cite{yongp3,yongp4,gan94}
\begin{equation}\label{QFT}
p_{\gamma}\equiv\frac{dN}{d\varepsilon_{\gamma}}=2.1\times10^{-6}\frac{(1-y^{2})^{\alpha}}{y},
\end{equation}
where $y = \varepsilon_{\gamma}/E_{max}$, $\alpha = 0.7319-0.5898\beta_i$,
$\varepsilon_{\gamma}$ is energy of emitting photon, $E_{max}$ is the energy
available in the center of mass of the colliding proton-neutron
pairs, $\beta_i$ is the initial velocity of
the proton in the proton-neutron center of mass frame.

\begin{figure}[th]
\centering
\includegraphics[width=0.5\textwidth]{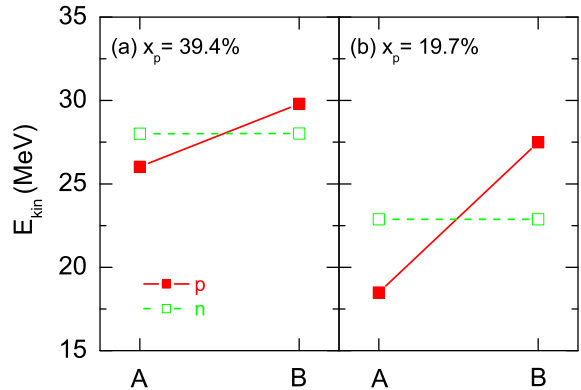}
\caption{ (Color online) Average kinetic energy of nucleon with different starting points of the HMT. The starting point of the HMT from their respective Fermi momentum is shown with A, while both start from the majority Fermi momentum is shown with B. Panel (a) is for the case of normal density nuclear matter with proton proportion $x_{p}$=39.4\% while panel (b) is for the case of normal density nuclear matter with proton proportion $x_{p}$=19.7\%. } \label{npkin}
\end{figure}
Since the starting momentum of the HMT of proton momentum distribution is different as shown in Fig.~\ref{npdis}, there should be a difference of the proton average kinetic energy with different starting momenta of the HMT of proton.
Fig.~\ref{npkin} shows the nucleon average kinetic energy changes with different starting momenta of the HMT. Because with case A and case B, the starting momentum of majority in the HMT is unchanged, one sees neutron kinetic energy almost keeps unchanged. While for proton, its average kinetic energy increases evidently, especially in large asymmetric matter. Compared panel (a) with panel (b), it is clearly shown that, due to the small number of correlated neutron-proton pairs in more asymmetric matter, the neutron average kinetic energy decreases evidently in more
neutron-rich matter.

In neutron star matter, the proton proportion $x_{p}$ is less than 10\%, it is thus deduced that the proton kinetic energy would have very large difference with different starting momenta in the HMT of proton. The large uncertainty of proton kinetic energy in neutron star matter would have evident influence on the dynamics of neutron stars \cite{mm2013}, such as the cooling of a Neutron Star, the superfluidity of protons, the isospin locking and the stiffness of the equation of state of the neutron stars, etc. Different distributions of the proton kinetic energy in nuclei would also affect the dynamics of nuclear reactions.

\begin{figure}[th]
\centering
\includegraphics[width=0.5\textwidth]{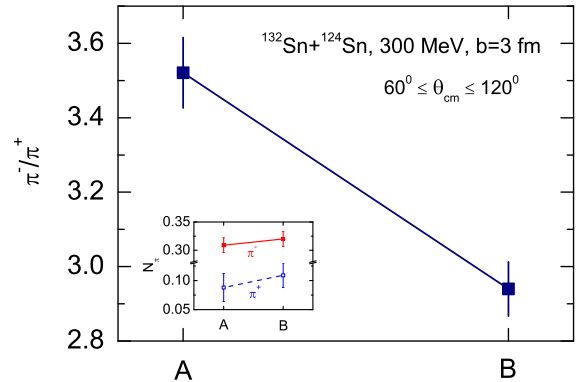}
\caption{ (Color online) The ratio of $\pi^{-}/\pi^{+}$ in $^{132}$Sn+$^{124}$Sn reactions at 300 MeV/nucleon incident beam energy with different proton starting momenta in the HMT.
$\theta_{cm}$ is polar angle relative to the incident beam direction. The inserted figure shows corresponding numbers of $\pi^{-}$ and $\pi^{+}$.} \label{pion}
\end{figure}
The change of proton average kinetic energy compared with neutron shown in Fig.~\ref{npkin} with different starting momenta of the $1/k^{4}$ momentum distribution causes an evident difference of proton and neutron average kinetic energies. The value $E_{kin}^{p}-E_{kin}^{n}$ changes from about -2 to 2 MeV in matter with $x_{p}$=39.4\% and the difference changes more evidently in more neutron-rich matter with $x_{p}$=19.7\%. The change of $E_{kin}^{p}-E_{kin}^{n}$ should affect the dynamics of pion production in heavy-ion collisions at intermediate energies, especially the value of the $\pi^{-}/\pi^{+}$ ratio in neutron-rich reactions. In heavy-ion collisions at intermediate energies, $\pi^{-}$ is mainly from neutron-neutron collision and $\pi^{+}$ is mainly from proton-proton collision. And more $\pi$'s are produced with the increase of nucleon-nucleon collision energy. Therefore, the evident change of $E_{kin}^{p}-E_{kin}^{n}$ with different starting points of the HMT should affect the $\pi^{-}/\pi^{+}$ ratio in neutron-rich reactions.

Since related pion measurements in $^{132}$Sn+$^{124}$Sn at 300 MeV/nucleon incident beam energy are ongoing at Radioactive Isotope Beam Facility (RIBF) at RIKEN in Japan \cite{exp1,exp2}, I thus use this reaction as an example to show how the $\pi^{-}/\pi^{+}$ ratio is affected by the starting point of the HMT in colliding nuclei. Fig.~\ref{pion} shows the ratio of $\pi^{-}/\pi^{+}$ in $^{132}$Sn+$^{124}$Sn reactions at 300 MeV/nucleon incident beam energy with different proton starting momenta in the HMT. As expected, there is a clear decrease of the value of $\pi^{-}/\pi^{+}$ ratio when changing the starting momentum of proton in the HMT from the proton Fermi momentum to the majority neutron Fermi momentum.
The effect reaches nearly 20\%. From the inserted figure, it is shown that such effect is mainly caused by the $\pi^{+}$ production. Therefore, the $\pi^{-}/\pi^{+}$ ratio in heavy-ion collisions at intermediate energies could be a probe of the starting momentum of the HMT of proton in neutron-rich matter.

\begin{figure}[th]
\centering
\includegraphics[height=9.0cm,width=0.45\textwidth]{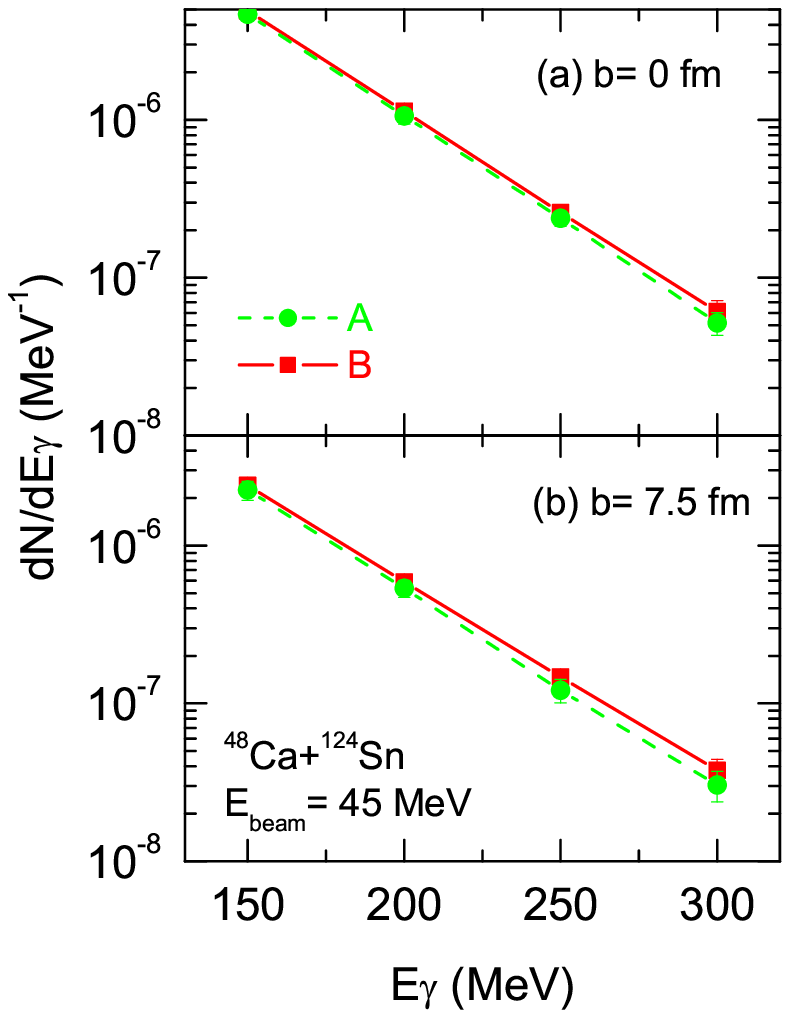}
\includegraphics[height=9.0cm,width=0.45\textwidth]{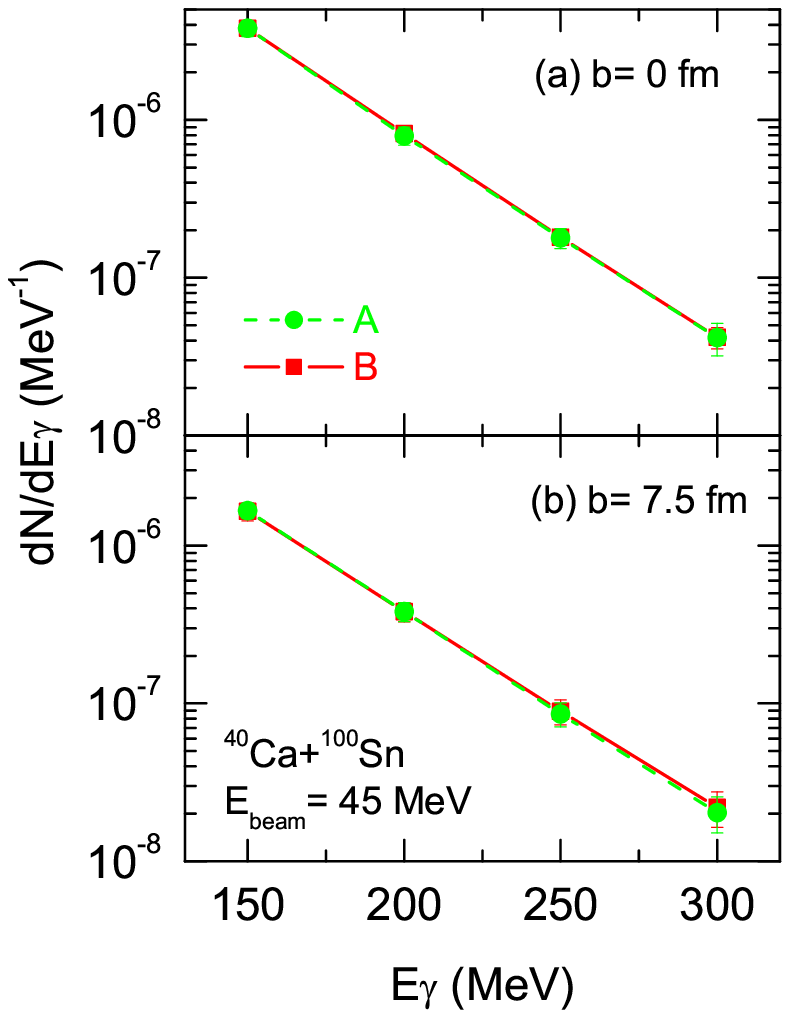}
\caption{(Color online) Top: Hard photon production in $^{48}$Ca + $^{124}$Sn reactions at 45 MeV/nucleon in central and peripheral collisions with different starting momenta of proton in the HMT. Bottom: Same as top panel, but for $^{40}$Ca + $^{100}$Sn.} \label{photo}
\end{figure}
\begin{figure}[th]
\centering
\includegraphics[width=0.5\textwidth]{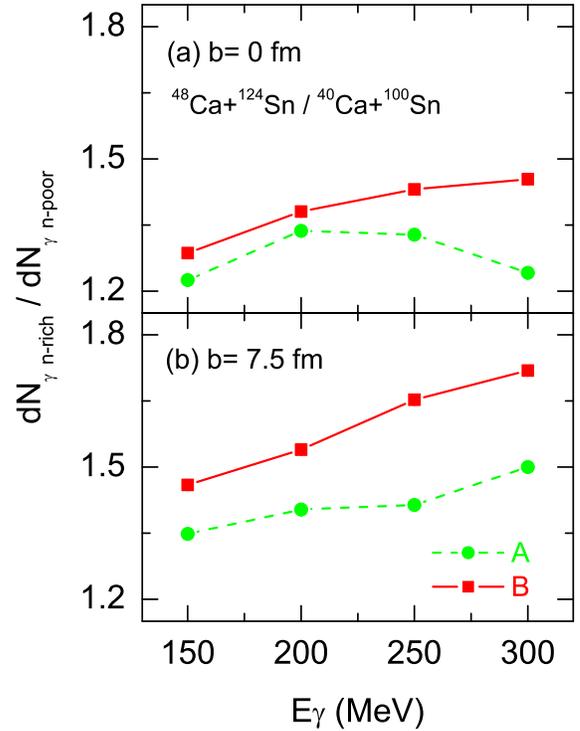}
\caption{ (Color online) The ratio of hard photon productions in neutron-rich and neutron-deficient reactions at incident beam energy of 45 MeV/nucleon in central and peripheral collisions with different starting momenta of proton in the HMT.} \label{photoratio}
\end{figure}
The hadron probe $\pi^{-}/\pi^{+}$ ratio may have stronger final-state interactions with other surrounding nuclear matter. While the electromagnetic probe, such as hard photon, has nearly no final-state interactions with other surrounding hadronic matter after it produces in heavy-ion collisions. I thus in this study try to see if the hard photon production can be used to probe the starting momentum of proton in the HMT. The top panel of Fig.~\ref{photo} shows the hard photon production in $^{48}$Ca + $^{124}$Sn reactions at 45 MeV/nucleon in central and peripheral collisions with different starting momenta of proton in the HMT. Because there is an increase of proton average kinetic energy when changing the starting momentum from the proton Fermi momentum to the majority neutron Fermi momentum, one sees more energetic photons are produced in $^{48}$Ca + $^{124}$Sn reactions at 45 MeV/nucleon, especially for peripheral collisions of neutron-rich nuclei. In more neutron-rich matter, the majority Fermi momentum becomes more larger than the minority. Thus changing the starting momentum of minority in the HMT to the majority's Fermi momentum would cause a larger increase of minority's kinetic energy. While for heavy-ion collisions with equal numbers of neutron and proton, it is expected that the effect of changing starting point in the HMT would disappear. The bottom panel of Fig.~\ref{photo} shows the hard photon production in $^{40}$Ca + $^{100}$Sn reactions at 45 MeV/nucleon in central and peripheral collisions with different starting momenta of proton in the HMT. It is clearly seen that changing the starting point of proton in the HMT has no effects on the energetic photon production.

To see more clearly the effects of different starting momenta of proton in the HMT on the energetic photon production, as shown in Fig.~\ref{photoratio}, I made a ratio of hard photon productions in neutron-rich and neutron-deficient reactions with different starting momenta of proton in the HMT. The ratio of energetic photon productions in neutron-rich and neutron-deficient reactions can not only reduce some theoretical systematic errors \cite{yongp2011}, but also clearly demonstrate the effects of different starting momenta of proton in the HMT on the energetic photon production. From Fig.~\ref{photoratio}, it is seen that the ratio of energetic photon production, especially in peripheral collisions, is sensitive to the choice of starting momentum of proton in the HMT.


In np-SRC dominance picture, it is in argument that how the transition momentum of minority from mean-field momentum distribution to correlated momentum distribution is determined in asymmetric nuclear matter. Based on the nuclear transport model, it is found that the $\pi^{-}/\pi^{+}$ ratio and the energetic photon production in neutron-rich nuclear reactions at intermediate energies can probe the proton transition momentum in neutron-rich matter. These studies may help us to understand the proton momentum distribution in asymmetric matter which have implications in both nuclear physics and astrophysics.


The author thanks Prof. Bao-An Li for useful communications.
The work is supported by the National Natural Science
Foundation of China under Grant Nos. 11375239, 11435014.

\end{document}